\documentclass[conference]{IEEEtran}
\usepackage{cite}
\usepackage{amsmath,amssymb,amsfonts}
\usepackage{algorithm}
\usepackage{algpseudocode}
\usepackage{graphicx}
\usepackage{textcomp}
\usepackage{xcolor}
\usepackage[hyphens]{url}

\def\BibTeX{{\rm B\kern-.05em{\sc i\kern-.025em b}\kern-.08em
    T\kern-.1667em\lower.7ex\hbox{E}\kern-.125emX}}

\title{DEAP: \underline{D}esign Space \underline{E}xploration for DNN \underline{A}ccelerator \underline{P}arallelism}
\author{\IEEEauthorblockN{Ekansh Agrawal}
\IEEEauthorblockA{
\textit{University of California, Berkeley}\\
agrawalekansh@berkeley.edu} 
\and
\IEEEauthorblockN{Xiangyu Sam Xu}
\IEEEauthorblockA{
\textit{University of California, Berkeley}\\
xiangyu.xu@berkeley.edu} 
}

\begin{document}
\maketitle
\thispagestyle{plain}
\pagestyle{plain}

%%%%%% -- PAPER CONTENT STARTS-- %%%%%%%%

\begin{abstract}

The boom in Large Language Models (LLMs) like GPT-4 and ChatGPT has marked a significant advancement in artificial intelligence. These models are becoming increasingly complex and powerful to train and serve. This growth in capabilities comes with a substantial increase in computational requirements, both in terms of hardware resources and energy consumption. The goal of this paper is to showcase how hardware and software co-design can come together and allow us to create customized hardware systems for specific LLM workloads. We propose a simulation workflow that allows us to combine model parallelism techniques with a multi-accelerator simulation framework for efficiency metrics. We focus on inference workloads and report power, cycle, and latency metrics upon performing a design space exploration search over multiple software and hardware configurations.

\end{abstract}

\section{Introduction}

Coupled with the growing size of LLMs is the ever-increasing cost of the computational and memory requirements to robustly serve these models. With the end of Dennard's Scaling, providing sufficient power to a single System-on-Chip (SoC) large enough to deliver the computational power required to complete a Large Language Model (LLM) computation has become challenging. The natural solution to this problem is to leverage the compute power of multiple accelerators instead of just one. However, this approach of hardware multiplexing introduces it's own inherent set of challenges. For starters, the communication bottlenecks between multiple accelerators can actually limit the potential speedups gained. Distributing the workload evenly across GPUs can be challenging, particularly when dealing with variable-sized inputs or residual connections. Fortunately, advancements such as Nvidia's NVLink \cite{nvlink} and Google's TPUv4 OCS technology \cite{jouppi2023tpu} have provided high-bandwith -- albeit expensive -- communication interface solutions to mitigate some of the problems. Still, simply splitting an execution trace amongst multiple-accelerators is not a cookie cutter approach to running a LLM faster.

In hyperscale data centers, where workloads can vary greatly, adaptability is crucial and is why FPGA accelerators are often employed. FPGAs can offer greater energy efficiency than GPUs which are often the de-facto accelerators of choice. FPGAs can be scaled more easily compared to other hardware solutions which is the main motivation behind this research project on design space exploration (DSE). DSE is crucial with FPGAs accelerators as it helps engineers optimize hardware configurations to meet performance, power, and resource constraints. We can leverage simulation to test different configurations off full-system hardware simulation platform that makes it easy to validate, profile, and debug RTL hardware \cite{karandikar2018firesim, parashar2019timeloop}. Given the architecture for a single accelerator, we aim describe a broad space of hardware topologies for multiple accelerators to be searched over by our DSE algorithms. In this research project, we will mostly use the Gemimini accelerator, a RoCC systolic array accelerator with non-standard RISC-V ISA \cite{genc2021gemmini}, and explore different inter-accelerator and intra-accelerator hardware configuration through DSE algorithms. 

Current research focuses on software/algorithm level exploration by setting the hardware setting as an invariant. This invariant is usually in the form of a single accelerator. We aim to explore how to incorporate hardware directly into the system design. By expressing a LLM architecture, we want to be able to use multi-accelerator simulation to perform an exhaustive search and find hardware configurations that maximize metrics for power, latency, and total cycles spent. Our proposed workflow is as followed:

\begin{enumerate}
  \item We use a hyperparameter search to create sequential LLMs of difference variations and sizes which allow us to simulate different LLMs. We also use a hyperparameters search to  different hardware configurations.
  \item We then break up the model architecture into sub-layers through different model parallelism techniques.
  \item We then use a scheduler to assign these sub-layer computations to different accelerators based on a few heuristics.
  \item We introduce DeapSim which simulates each accelerator's workload with Timeloop and simulates the accelerator interconnect to get metrics on power, cycles, and latency on the given workload.
\end{enumerate}

\section{Background}
\subsection{LLMs}

LLMs are primarily built upon the Transformer architecture, leveraging techniques like word embeddings and attention mechanisms to process large amounts of data effectively. They have capabilities ranging from understanding and generating text based on context to passing standardized tests and recognizing humor \cite{zhao2023survey}. Google's BERT was the first LLM to leverage the attention mechanism to construct first deeply bidirectional, unsupervised language representation, pre-trained using only a plain text corpus. BERT operates on two main training strategies: Masked Language Modeling (MLM) and Next Sentence Prediction (NSP) \cite{bert}.

The GPT (Generative Pretrained Transformer) series, developed by OpenAI represents a progression in the field of of LLMs with each iteration introducing significant advancements. The original GPT model, with its architecture based on the Transformer model introduced by in 2017, had 117 million parameters. It primarily used a stack of decoder blocks from the Transformer architecture. The model was trained to predict the next word in a sentence, learning to generate coherent text over time \cite{gpt1}. GPT-2 expanded significantly on this architecture, boasting 1.5 billion parameters. While it maintained the fundamental architecture of GPT-1, the increase in parameters allowed for more depth and complexity in learning patterns and language understanding. GPT-2's larger scale improved its ability to generate more coherent and contextually accurate text, demonstrating a significant leap in language modeling capabilities \cite{gpt2}. With GPT-3, the architecture underwent a massive scale-up, featuring an unprecedented 175 billion parameters. Although the fundamental architecture remained similar to GPT-2, focusing on the Transformer's decoder blocks, the sheer increase in size allowed GPT-3 to perform a wide range of language tasks with minimal task-specific training data \cite{gpt3}. As of April 2023, GPT-4 represents the latest iteration with further advancements in scale and complexity. \cite{gpt4}. Unofficial leaks claim the model have 1.75 trillion parameters coupled with 8 multi-agent mixture of experts architecture \cite{gpt4, moe, gp4leak}.

Meta's LLaMA (Large Language Model Meta AI) was introduced as a foundational language model with several versions varying in size: 7B, 13B, 33B, and 65B parameters. The model's training involved 1.4 trillion tokens for the 65B and 33B versions, and 1 trillion tokens for the 7B model \cite{llama}. LLaMA-2, an extension of the original LLaMA model, comes in a range of sizes from 7 billion to 70 billion parameters. It's an auto-regressive language model optimized for transformer architecture. LLaMA-2 utilizes supervised fine-tuning (SFT) and reinforcement learning with human feedback (RLHF) to align with human preferences for helpfulness and safety \cite{llama2}.

Combing these foundation models with technologies like retrieval-augmented generation opens up endless possibilities for creating AI solutions. \cite{rag}. But it's evident that these feats have only been possible due to the sheer size of their neural networks and their abilities to generalize vast amounts of data.

\begin{figure}[htp]
    \centering
    \includegraphics[width=9cm]{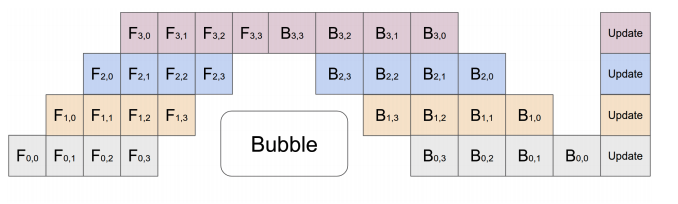}
    \caption{An example of a pipeline parallelism is shown here where we can split the model up into sections of execution and massively increase the throughput of inference.}
    \label{fig:pipeline}
\end{figure}

\begin{figure}[htp]
    \centering
    \includegraphics[width=8cm]{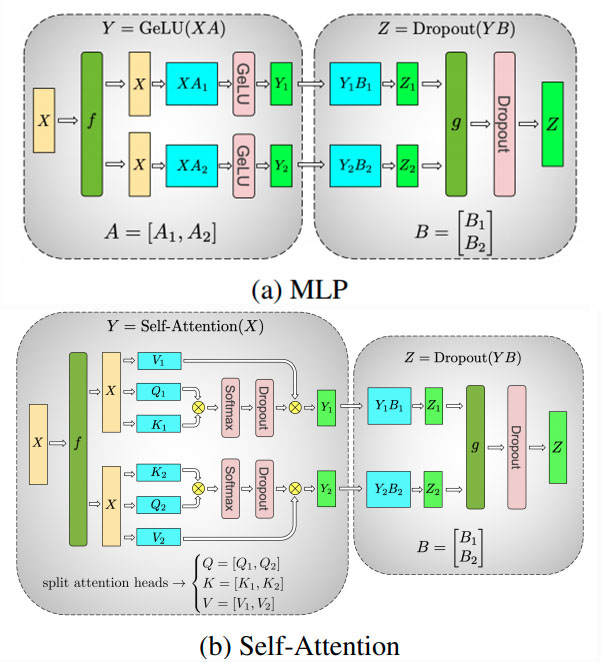}
    \caption{An example of tensor parallelism is shown above where we can decompose a larger computation amongst multiple accelerators and then compose the values back together to simulate the same giant calculation.}
    \label{fig:tensor}
\end{figure}

\subsection{Model Parallelism}

Due to the sheer sizes of these LLM, single-GPU training and inference is impractical. Model parallelism allows systems to leverage multi-accelerators and split up model execution across multiple devices. This allows us to run inference on models that are too large to fit onto one GPU.

A novel contribution in this area is an algorithm designed to optimize micro-batch size for efficient pipelining in multi-GPU environments. This algorithm, by reducing the overhead involved in determining the optimal micro-batch size, takes into account factors like the number of GPUs and their memory capacities. Notably, it has been successfully applied to U-Net, a deep neural network used in medical imaging, demonstrating improvements in image throughput and mini-batch size capacity. Additionally, the study explores the impact of normalization techniques like batch and group normalization in distributed deep learning settings, noting how the latter can mitigate performance degradation issues often seen with batch normalization in distributed environments \cite{choi2023towards}.

Parallelism in deep learning models can be implemented in various ways, such as Sharded Data Parallelism (Zero-DP) \cite{rajbhandari2020zero} and Naive Model Parallelism (Vertical). In Sharded DDP, a model is divided across multiple GPUs, with each GPU handling a fraction of the model's parameters and a subset of the input data \cite{zhao2023pytorch}. On the other hand, Naive Model Parallelism involves distributing different groups of model layers across multiple GPUs. This method is termed 'vertical' as it effectively slices the model layers vertically across the GPUs. For instance, in an 8-layer model, layers 0-3 might be placed on GPU0 and layers 4-7 on GPU1. As data travels through the layers, it switches between the GPUs correspondingly.

Pipeline Parallelism, as shown in Figure  Fig.\ref{fig:pipeline}, extends the idea of model parallelism but spliting sequential layers further into mini-blocks. A simple partition algorithm aligns the network into $K$ cells and places the $k$-th cell on the $k$-th accelerator. We stagger the computation of the mini-blocks to allow for greater throughput in both inference and training settings \cite{huang2019gpipe}. We prefer pipeline parallelism due to it's low overheard in terms of implementation. Tensor Parallelism, as shown in Figure  Fig.\ref{fig:tensor}, breaks up a computation into intermediaries which can be multiplexed and then accumulated. We employ a hierarchical layer-wise dynamic programming method to search for the partition for each layer \cite{song2019hypar}.

\subsection{ML Systems for LLMs}

Significant advancements in model architectures have contributed to more efficient LLMs. Techniques like pruning \cite{prune}, quantization \cite{quantization}, and distillation \cite{distillation} have been employed to reduce the computational load. ML systems however serve as the backbone for deploying, managing, and optimizing LLMs in production, making them a critical component for leveraging these language models effectively in various applications.

The use of specialized hardware for domain specific applications like GPUs and TPUs has greatly accelerated ML inference. Google TPUv4, built with optical circuit switches and systolic array architecture, boasts a 275 teraflops int8 performance with a memory bandwith of 900 GB/s \cite{jouppi2023tpu}. Nvidia's flagship A100 general-purpose GPU with CUDA and specialized tensor cores delivers 312 teraflops int8 performance with 2 TB/s memory bandwith \cite{nvidia}. 

Libraries such as TensorFlow \cite{tensorflow}, PyTorch \cite{pytorch}, and JAX \cite{jax} have continuously evolved, providing optimized algorithms that make better use of underlying hardware. These improvements include better memory management, optimized tensor operations, and support for asynchronous computation. Compiling efficient kernels with a JIT compiler generates efficient kernel can decrease the memory transfer burden of accelerators \cite{torchy}.

Distributed computing techniques have also enabled the parallel processing of large-scale ML tasks. Statistical model multiplexing has proven to reduce the average completion time of bursty requests \cite{alphaserve}. Iteration-level scheduling that schedules execution at the granularity of iteration has shown to improve throughput by 36.9× with batched request \cite{orca}.

\subsection{Design Space Exploration}

Hardware-software co-design in recent years has become a pivotal area in optimizing the performance of specialized hardware, particularly for Deep Neural Networks (DNNs). This interdisciplinary field integrates both hardware design and software development, focusing on how these two aspects can be mutually optimized for better performance, energy efficiency, and cost-effectiveness.

The process of hardware design space exploration (DSE)\cite{DSE} is notably complex and resource-intensive. It involves examining various hardware design parameters and software mappings to enhance application performance. The primary objectives in this field are twofold: mapping search and hardware search. Mapping search aims to find effective ways to utilize hardware resources for high-performance computing, while hardware search strives to achieve balanced design goals, such as minimizing energy-delay or area-delay products.

In the context of DNNs, the complexity of mapping has led to the development of specialized DNN compilers and accelerator-aware mapping techniques. These tools and methods are designed to efficiently map neural network computations to specific hardware architectures, considering factors like parallelism, memory hierarchy, and computational capabilities.

The exploration of hardware parameters is another critical aspect, where researchers focus on finding the optimal hardware configurations that meet the desired objectives of performance and efficiency\cite{Hardware_Search}. This involves a meticulous process of testing and evaluating different hardware designs under various conditions and constraints.

Recent studies\cite{Learn_DNN} have introduced co-exploration frameworks that address both mapping and hardware designs simultaneously, aiming for higher efficiency and reduced development costs. These frameworks typically employ a two-loop process. The first loop involves sampling a hardware design and then finding high-performance mappings for it. The second loop uses the best mapping to guide further hardware optimization. Optimization techniques in these frameworks range from heuristics and black-box optimization (which includes methods like genetic algorithms and reinforcement learning) to white-box optimization (using mathematical models and techniques like gradient descent).

However, the two-loop approach can be prone to combinatorial explosion due to the vast search space. To mitigate this, single-loop searchers like DiGamma and Interstellar have been proposed. These methods focus on finding high-performance mappings first and then deducing the minimal hardware requirements \cite{kao2022digamma, interstellar}. This approach reduces the size of the search space but may have limitations in exploring the full potential of hardware-mapping combinations. A notable advancement in this area is the Differentiable Model-Based One-Loop Search (DOSA)\cite{DoSA}. DOSA uses a differentiable white-box model for the analytical performance and energy model. By employing gradient descent, it optimizes mapping variables efficiently, thus enabling the exploration of a comprehensive set of mappings and hardware configurations without extensive reliance on simulators.

The current focus in hardware-software co-design, particularly for DNN accelerators, has been primarily on DSE for single-accelerator systems. This approach has led to significant advancements in optimizing individual accelerator performance. However, as computational demands increase, there is a growing need to extend these DSE methodologies to multiple accelerators. This transition represents a shift towards harnessing the combined computational power of multiple specialized processors. Adapting existing DSE techniques to a multi-accelerator context involves addressing new challenges like inter-accelerator coordination and workload distribution. By leveraging and scaling these techniques, research in multi-accelerator DSE can lead to more powerful and efficient computing systems, opening new frontiers in high-performance computing and AI applications.

\subsection{Multi-Accelerator Simulation}

In the realm of DSE for DNN accelerators, the simulation tool plays a critical role. It's the backbone of the exploration process, used to evaluate each design point and facilitate efficient search. Typically, DSE for DNN accelerators relies on pre-RTL software simulation tools, which offer the speed necessary for effective exploration.

The Roofline \cite{Roofline} model has set a standard for software simulation, offering a clear framework for assessing peak performance and memory bandwidth constraints of a system. Building on this, Timeloop \cite{Timeloop}, in particular, stands out in the DSE landscape for DNN accelerators. It offers a comprehensive and flexible way to describe the key attributes of various DNN architectures and their implementation features. This description serves as the input for a fast and accurate analytical model. Timeloop's strength lies in its ability to accommodate a broad spectrum of architectures, allowing for extensive exploration within a unified framework.

Furthermore, Timeloop integrates this architectural exploration with a sophisticated mapping tool. This tool is designed to identify optimal mappings of any given workload on the targeted architecture. Such a feature is crucial for making fair comparisons between different architectures. It brings a level of systematic rigor to the DNN accelerator design process, transforming it from an art to a more structured and methodical practice. It excels in modeling and exploring the design space of individual accelerators, providing detailed insights into their performance and efficiency. Timeloop's capabilities are particularly tailored to the unique requirements of single-accelerator systems, making it an invaluable resource in optimizing these architectures.

In contrast, Astra-Sim\cite{AstraSim} has emerged as a new and promising tool, focusing on the complexities of multi-accelerator systems, particularly in distributed training scenarios. This tool is geared towards understanding and optimizing the interactions and data distributions across multiple accelerators. Astra-Sim's development marks a significant step towards addressing the needs of more complex, distributed computing environments. However, it currently faces challenges in scalability and versatility. This limits its ability to support diverse parallelism strategies, network architectures, and memory models, reflecting the broader challenges in adequately simulating the multifaceted nature of multi-accelerator systems.

\section{Creating a Workload}
\subsection{LLM Architecture}
In order to define a ML workload to simulate, we must first define a model representation. Due to the parallelism techniques explored in this paper, the only constraint on our model expression is that we must craft a sequential model. This ensures that each stage of the model processes the data in a specific order. This maintains the integrity of the model's forward pass. We draw inspiration from GPT-2 transformer encoder--decoder stack and vary different hyperparameters to construct unique LLMs from this base configuration. Figure 4 shows the construction of the single transformer encoder-decoder layer. Table  Fig.\ref{table:parameters} shows the different hyperparameters we can tune in order to get different LLM workloads. Though though the search space for model architecture seems sparse, varying these hyperparameters gives us enough varied tensor workloads to simulate within Timeloop.

\begin{scriptsize}
\begin{table}[h!]
  \centering
  \caption{Hyperparameters for generating LLMs}
  \label{table:formatting}
  \begin{tabular}{|l|l|}
    \hline
    \textbf{Hyperparameter} & \textbf{Value}\\
    \hline
    --embedding-dimension & Positional Encoding Dimension \\
    \hline
    --forward-dimension & Feedforward Transformer Block Dimension \\
    \hline
    --num-heads & Number of Heads in Attention Block \\
    \hline
    --num-decoder-layers & Number of Stacked Decoder Layers \\
    \hline
    --vocab-size & Number of Total Tokens \\
    \hline
  \end{tabular}
\caption{Above we show the parameters that can be tuned in order to generate LLM architectures of different sizes.}
\label{table:parameters}
\end{table}
\end{scriptsize}

\begin{figure}[htp]
    \centering
    \includegraphics[width=7cm]{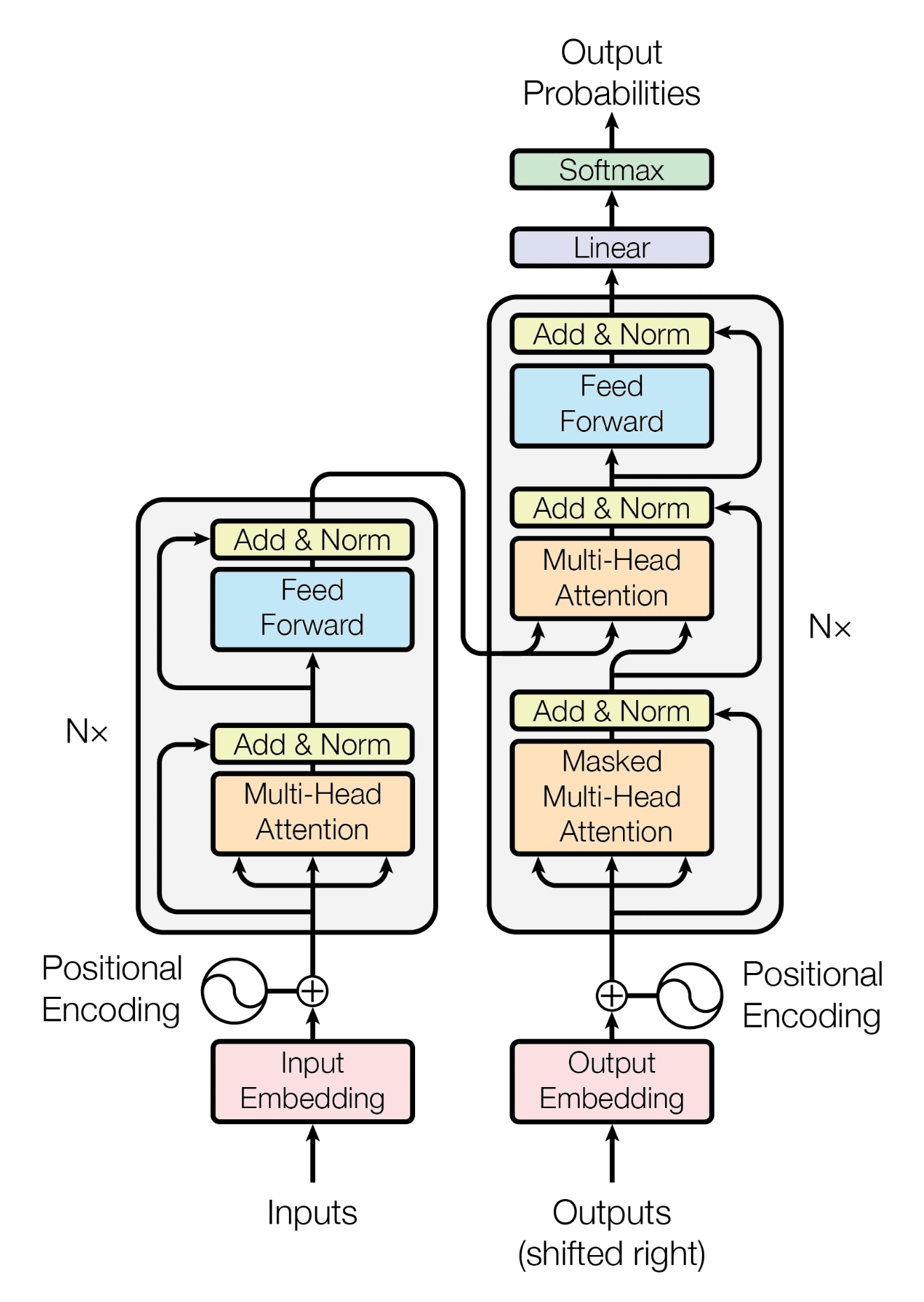}
    \caption{Shown above is the representation for a single transformer encoder-decoder layer. In addition to the various hyperparameters for the internal feed-forward layers, can vary the number of decoder and encoder layers to create LLMs of difference sizes for our workload.}
    \label{fig:galaxy}
\end{figure}

\subsection{Model Parallelism}

Now that all layers of the LLM have been initialized, we combine tensor parallelism with pipeline parallelism to split the execution graph. Since most of the computations within the transformer layers are express as feed forward layers, we split each layer with tensor parallelism. The layers themselves are then split based on the number of microbatches which is expressed as a hyperparameters. Since our model is sequential, these batches are fed sequentially through the different stages of the model. While one batch is being processed in one stage, another batch can be processed in a different stage.

\subsection{Graph Scheduling}

\begin{algorithm}
\caption{Subgraph Scheduling Algorithm}
\begin{algorithmic}[1]
\State $tasks \gets \text{list of all sublayers to be scheduled}$
\State Set current time steps for each worker to zero
\State Initialize a set for tracking finished sub-layers

\While{not all tasks are finished}
    \For{each worker}
        \If{no task is currently assigned to the worker}
            \State Set flag indicating task assignment to false
            \For{each task in the list of tasks}
                \If{all dependencies of layer finished}
                    \State Assign the task to the current worker
                    \State Remove the task from the list
                    \State Set flag for task assignment to true
                \EndIf
            \EndFor
            \If{no task was assigned}
                \State Assign a NOP to the worker
            \EndIf
        \EndIf
        \State Increment the worker's current time step
    \EndFor
\EndWhile

\State \Return the final workload distribution for all workers
\end{algorithmic}
\end{algorithm}

Once all the sub-layers have been defined, we remove any layers that are not used in the forward inference like batch norm and dropout layers. We then use a scheduling algorithm to determine which accelerator runs which sub layers at a given time step. We consider each sub-layer to be a task and we consider the duration of each task to be the number of floating point operations (FLOPs) it takes for the task to finish executing. Since most of computations in our LLMs can be expressed as a combination of matrix multiplications, we can use the the following equation the calculate the FLOPs of a sub-layer.

\begin{equation}
    \text{FLOPs} = (2I - 1) \times O
\end{equation}
where \( I \) is the number of input features, and \( O \) is the number of output features.

We then use Algorithm 1 to to schedule the sub-layers amongst the N accelerators and assign NOPs to the accelerators at certain time steps where computation is not being performed. Upon the termination of the scheduling algorithm, we output a scheduling workload that consists a a nested list of dictionaries. Each index in the outer list represent the schedule assignment for accelerator $i$. Each element at time step $j$ for an accelerator $i$ represents a dictionary of parameters that can be fed into Timeloop for simulation. We have an example payload defined below for a feed forward layer with a batch size of 32 and an output feature dimension of 2500 for accelerator 0 and timestep 0:

\begin{verbatim}
{
    'C': 32,
    'Hdilation': 1,
    'Hstride': 1,
    'K': 32,
    'N': 1,
    'P': 2500,
    'Q': 1,
    'R': 1,
    'S': 1,
    'Wdilation': 1,
    'Wstride': 1,
    'type': 'LinearLayer'
}
\end{verbatim}

Where for a given layer, $R$ is weight width, $S$ is weight height, $P$ is output width, $Q$ is output height, $W$ is input width, $H$ is input height, $C$ is input channel size, $K$ is output channel size, and $N$ is batch size. The dilation and stride parameters are usually set for convolution operations, so we default them to 1 since our LLMs do not use any convolutional parameters in their forward pass.

\section{DeapSim}
We propose DeapSim for our multi-accelerator simulation, which represents an innovative leap in the simulation of distributed deep learning systems, building upon the foundational work of Timeloop\cite{Timeloop} and extending the Roofline\cite{Roofline} model to a broader context. This simulation platform is intricately designed to be cognizant of both software scheduling intricacies and hardware configurations, with a special emphasis on the topology of chip groups. By integrating these elements, DeapSim offers a comprehensive tool for exploring and optimizing the complex interplay between diverse accelerators within a distributed network. It stands as a testament to advanced co-design methodologies, enabling the detailed examination and fine-tuning of system-wide parameters that influence the efficiency and efficacy of deep learning training at scale.

\begin{figure*}[htp]
    \centering
    \includegraphics[width=18cm]{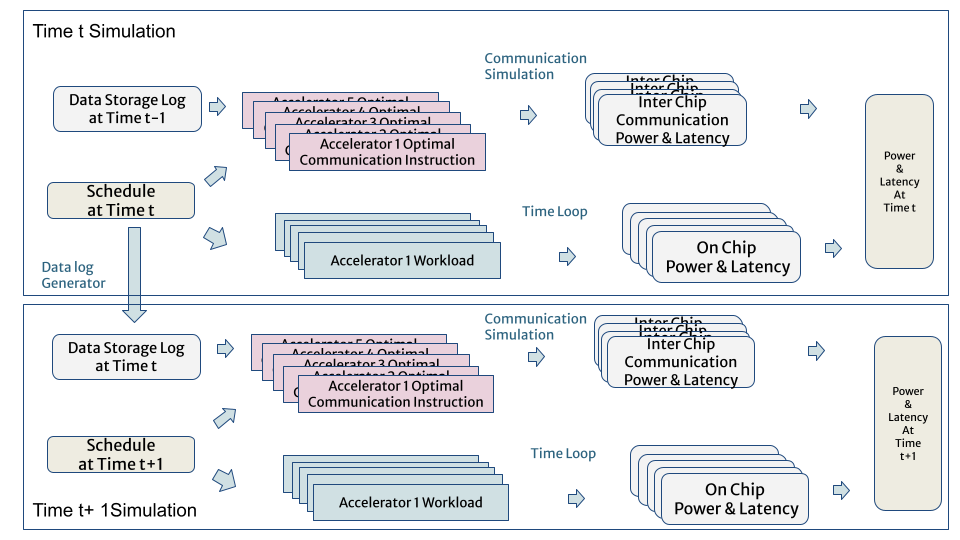}
    \caption{At each time step, the simulation uses logs from the previous step and the current schedule to determine workloads for accelerators. It then concurrently simulates on-chip processing and inter-chip communication, evaluating the system's power and latency. These simulations inform adjustments for the next time step, facilitating an iterative optimization process.}
    \label{fig:DeapSim}
\end{figure*}

\subsection{Hardware Architecture}
\begin{figure}[htp]
    \centering
    \includegraphics[width=5cm]{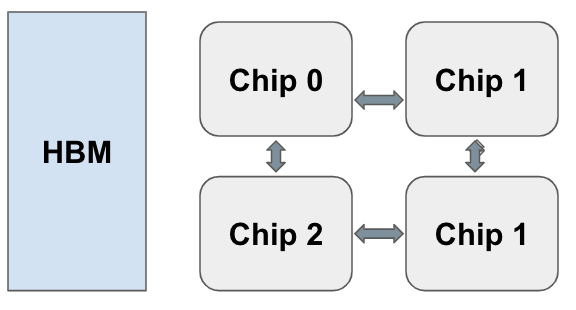}
    \caption{An example of a simple 2D-Torus topology that connects 4 accelerators together.}
    \label{fig:2d-torus}
\end{figure}

\begin{figure}[htp]
    \centering
    \includegraphics[width=8cm]{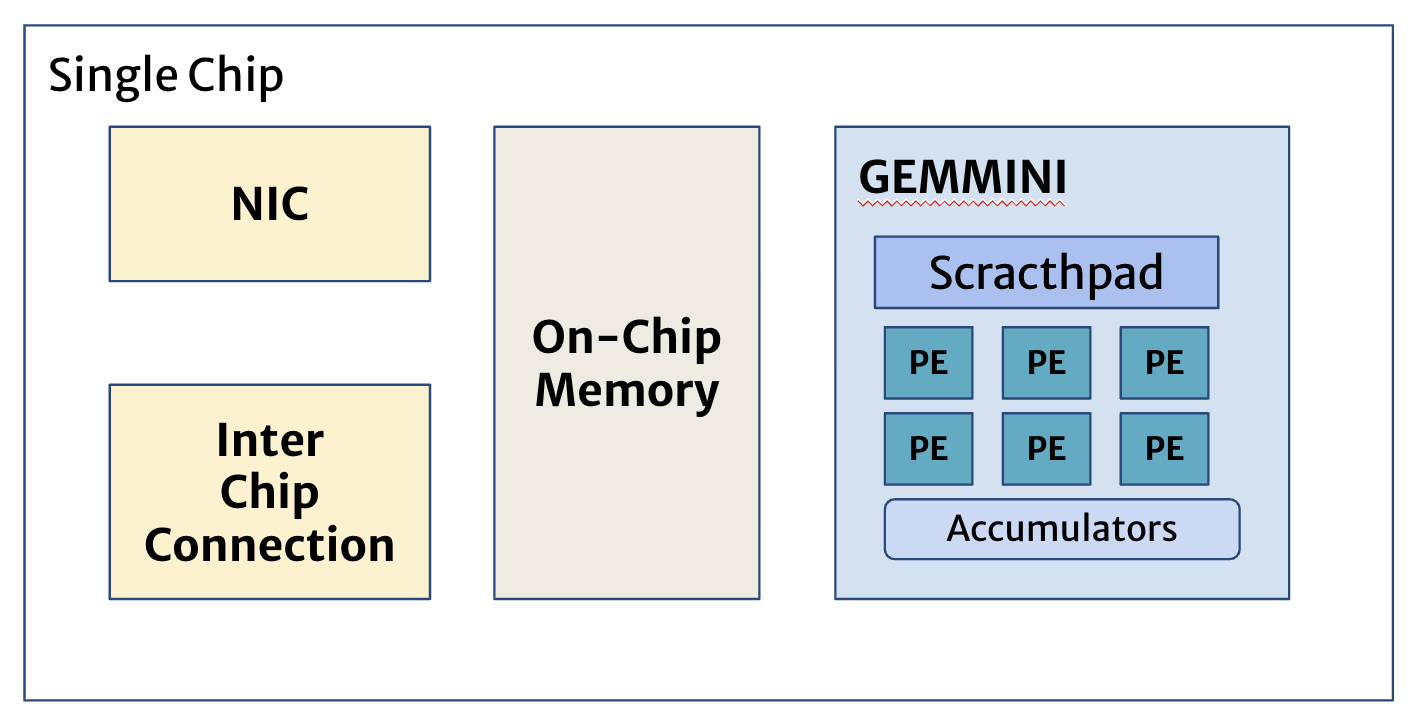}
    \caption{An example of a configuration for a single Gemmini accelerator.}
    \label{fig:single_acclerator}
\end{figure}

DeapSim is architected to simulate a multi-chip environment with a base configuration that includes High Bandwidth Memory (HBM) \cite{HBM} and a customizable number of interconnected chips. Each chip features an on-chip memory, a Network Interface Controller (NIC) that maintains connections with other chips and the HBM, and an inter-chip connection enabling direct communication between chips. Every chip houses a single accelerator \cite{Gemmini} whose configuration is user-defined, as shown in Figure \ref{fig:single_acclerator}, allowing for specificity in simulation.

The topology that outlines the inter-chip connections is designed for scalability and can be tailored by the user. It allows for definition of the interconnect technology and the physical distances between chips, supporting a range of common topologies such as 2D Torus, 3D Torus, and 1D pairs. This flexible setup in DeapSim enables users to explore various chip arrangements and communication strategies, facilitating a comprehensive analysis of different hardware configurations and their implications on system performance. There is a sample topology description for 2$\times$2 2D-torus topology :
\begin{verbatim}
{
  Topology: 2D-torus
  Size: 4
  Dimension: 2
};
{
  Chip_id: 0
  Connected_chip_id: {1, 3}
  Connected_chip_distance: {1, 1}
}
{
  Chip_id: 1
  Connected_chip_id: {0, 2}
  Connected_chip_distance: {1, 1}
}
{
  Chip_id: 2
  Connected_chip_id: {1, 3}
  Connected_chip_distance: {1, 1}
}
{
  Chip_id: 3
  Connected_chip_id: {0, 2}
  Connected_chip_distance: {1, 1}
}
\end{verbatim}

\subsection{Software Mapping Configuration}
DeapSim uses an innovative approach to software mapping through its utilization of a generalized schedule format. This format comprehensively details the distribution and execution strategy of workloads across multiple accelerators. In the schedule file, DeapSim specifies how a larger workload is divided into smaller, manageable sub-tasks. Crucially, it outlines which sub-workload is allocated to a specific accelerator and the precise timing for each task's execution.

This method of scheduling is a key strength of DeapSim, as it accommodates a wide array of parallelism strategies. By detailing the distribution of sub-workloads across different accelerators and their execution timelines, DeapSim ensures a highly efficient and optimized utilization of resources. This generalized scheduling approach enables DeapSim to adapt flexibly to various computing architectures and parallel processing techniques, making it a versatile and powerful tool in the realm of multi-accelerator systems.

\subsection{Communication Stage Simulation}
At each communication stage, DeapSim directs the on-chip memory of each chip to store the output from the preceding processing stage and to load the input for the subsequent one. Since the software schedule lacks explicit instructions for data movement, DeapSim deduces the optimal communication paths.

DeapSim maintains a data log at each time step, detailing the storage status of each chip's on-chip memory per the schedule. Leveraging this log, DeapSim ascertains the most effective communication flow for each stage. It reviews the assigned workload for every accelerator, cross-references the storage log to locate required data, and determines the data's current location. If the needed data is not on any chip, it is sourced from the HBM. If the needed data is on the requesting chip, no communication will happen. When another single chip hold the needed data, a direct transfer from that chip to the requesting chip occurs. When multiple chips have the data, DeapSim evaluates the chip group's connection topology to facilitate the most efficient data exchange.

\begin{scriptsize}
\begin{table}[h!]
  \centering
  \caption{Inter Chip InnerConnection Configuration} 
  \label{table:formatting}
  \begin{tabular}{|l|l|}
    \hline
    \textbf{Number of Links} & \textbf{Latency (GB/s)}\\
    \hline
    NIC & 1 \\
    \hline
    1 & 180 \\
    \hline
    3 & 64 \\
    \hline
   12 & 25 \\
    \hline
  \end{tabular}
\caption{Above we show the fabric latency values that are used in our framework to simulate multi-accelerator latency.}
\end{table}
\end{scriptsize}

After extracting the dataflow, DeapSim will estimate the latency each communication by the following equations.
\begin{equation}
\text{Latency} = \text{technology factor} \times \text{data size} \times \text{distance}
\end{equation}

Notice that the technology factor here are defined by the topology, which is affected by the number of links on each chip. DeapSim uses the default technology factor from Google TPU v4\cite{jouppi2023tpu}. Meanwhile DeapSim make each connection have the constant power consumption.

The overall communication latency also include the memory bandwidth. Then, the overall communication stage latency on each single chip is:
\begin{align*}
\text{Latency}_{\text{chip c}} = &\max_{i \in \{\text{all communication on chip} \}}\text{Latency}(i) \\
&+ \frac{\sum_\_{i \in \{\text{all communication on chip} \}} i
}{\text{On-Chip Memory Bandwidth}}
\end{align*}

\subsection{Process Stage Simulation}
During each phase of the processing cycle, all accelerators within the system concurrently process their allocated data as defined by the software schedule. DeapSim meticulously executes this stage by utilizing the schedule to assign workloads to each accelerator. Following this, DeapSim employs Timeloop \cite{Timeloop} to estimate the latency and power consumption for each individual accelerator.

\subsection{Metric Evaluation}

With the individual latency and power metrics for each chip at every time step now available, we can calculate the aggregate latency and power consumption using the principles of synchronous training. Synchronous training is crucial in distributed deep learning as it orchestrates the operations on multiple accelerators to work in unison on the same data iteration. This harmonization not only bolsters efficiency and model accuracy but also prevents the discrepancies in data processing that are typical of asynchronous methods. By ensuring all accelerators update the model concurrently, synchronous training promotes uniform model improvements and accelerates convergence.

To quantify the overall latency at a given time $t$, we consider the chip experiencing the maximum communication and processing delays within that time step as shown in Figure  Fig.\ref{fig:DeapSim}, as these will dictate the cycle's duration:

\begin{figure*}[htp]
    \centering
    \includegraphics[width=18cm]{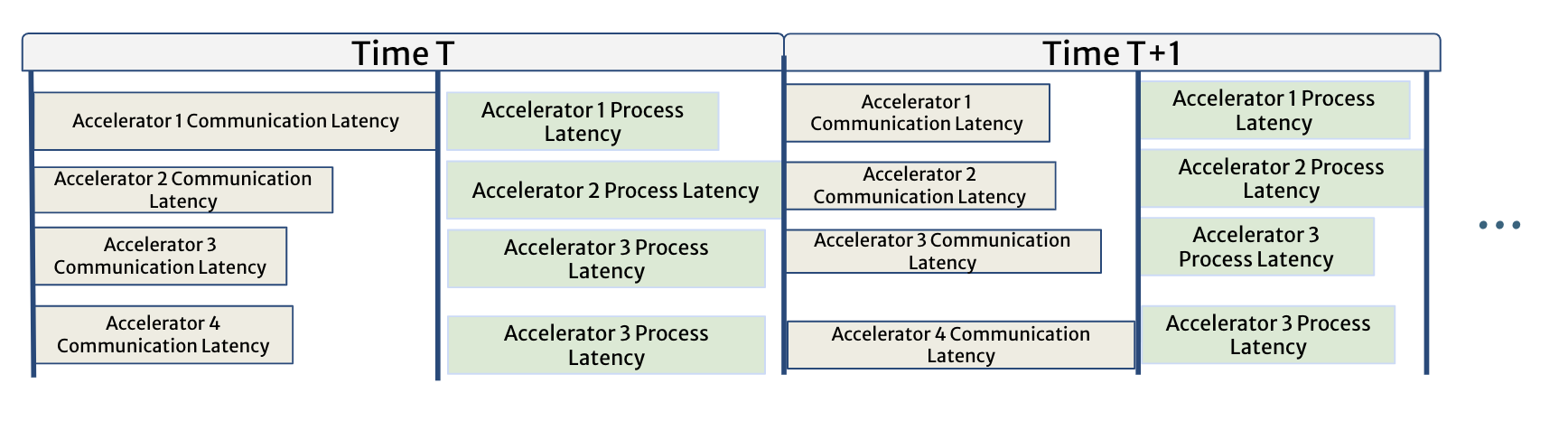}
    \caption{Above is a graphic showing the communication synchronization between accelerators. We block on all accelerators before starting running the simultation for a computation.}
    \label{fig:DeapSim}
\end{figure*}

\begin{align*}
\text{Overall Latency}_t = \max &\left( \text{Comm Latency all chips} \right) \\
&+ \max \left( \text{Process Latency all chips} \right)
\end{align*}

Similarly, the total power consumption at time $t$ is the sum of the power used for communication and processing across all chips:

\begin{align*}
\text{Overall Power}_t = \sum_{i=1}^{N} P_{\text{comm},i} + \sum_{j=1}^{N} P_{\text{proc},j}
\end{align*}

Finally, by summing these metrics across all time steps, we obtain the cumulative latency and power for the entire training operation, which provides a comprehensive view of the system's performance over the duration of the model training.

\section{Design Space Exploration}

DeapSim advances the field of hardware-software co-design by addressing the complex challenge of design space exploration (DSE) for distributed DNN accelerators. This comprehensive approach encompasses a meta configuration that includes the optimization of software schedules across multiple chips, determining the most effective workload distribution and timing to maximize performance. Alongside this, the hardware configuration is examined to decide the optimal number of chips, their interconnectivity, and the individual chip configurations, such as cache size and processing element count.

\begin{figure}[htp]
    \centering
    \includegraphics[width=9cm]{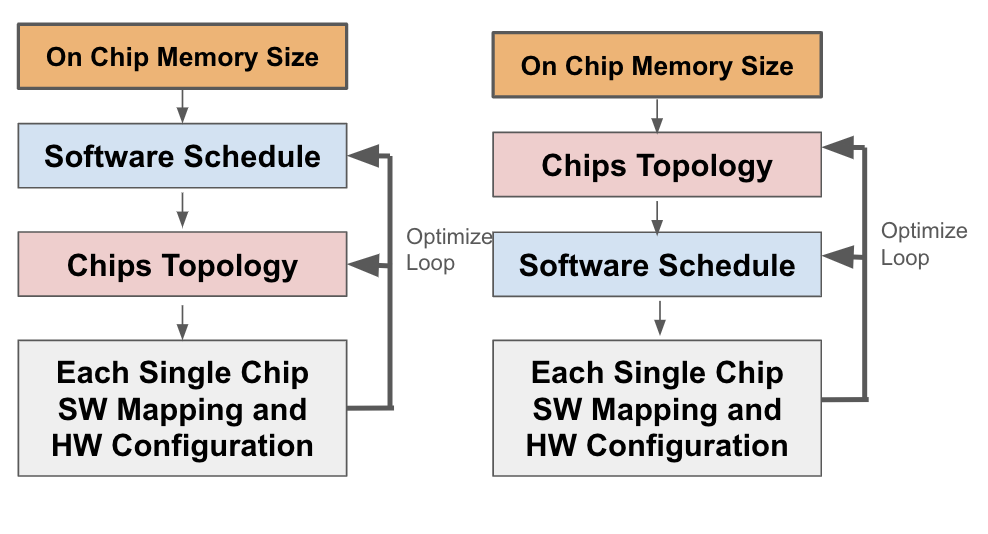}
    \caption{Two potential optimization flows within DeapSim for co-designing hardware and software in a multi-chip DNN accelerator environment. The left flow starts by setting the on-chip memory size, which informs the software scheduling decisions for workload distribution. This is followed by defining the chips' topology, and finally, refining the software mapping and hardware configuration for each chip. The right flow prioritizes chips topology before software scheduling. Both flows converge at a loop point where iterative optimization is applied to refine the overall system configuration.}
    \label{fig:dual_search}
\end{figure}

We propose a dual search flow within DeapSim as shown in Fig.\ref{fig:dual_search}. The process begins by determining the on-chip memory size, which guides the software schedule in effectively partitioning large workloads into smaller, manageable segments. Subsequently, the search can prioritize either the software schedule or the hardware configuration. Each layer of this process can employ search algorithms to iteratively refine and optimize the system's overall performance. This methodology not only elevates the efficiency of individual accelerators but also harmonizes their collective operation within a multi-accelerator framework.

This framework presents a generalized approach to model parallelism and topology optimization for distributed deep learning systems. It abstracts and adapts to any model parallelism strategy by allowing flexible software scheduling, which determines how a model is partitioned and executed across multiple DNN accelerators. This flexibility enables it to accommodate different parallelism paradigms, from data and model to pipeline parallelism.

In terms of topology, the framework is not restricted to any specific inter-chip connection schema. Instead, it supports a variety of topologies, from simple ones like 1D pairs to complex structures like 2D or 3D Torus configurations. This versatility allows for exploration of the most efficient paths for data movement and communication between chips, which is critical for optimizing the performance of large-scale, distributed neural network training.

By generalizing the considerations for both model parallelism strategies and chip topologies, this framework stands as a robust tool for DSE, facilitating the exploration of a vast design space to identify optimal configurations for a given set of hardware and software constraints.

\section{Evaluation}

\subsection{Hardware Topology Search Case}
In this section, we delve into the search for the most effective hardware topology tailored to various workloads, utilizing a predetermined software schedule as outlined in Section III. Our simulations are grounded in Gemmini, selected as the on-chip accelerator of choice\cite{Gemmini}. To model the inter-chip communications, we adopt configurations akin to those found in the TPU v4, and we calibrate Gemmini's operational frequency to 700MHz for consistency across simulations. Meanwhile, we set the on-chip memory to 32MB.

During each iteration of our optimization process, we implement a comprehensive search strategy that involves deploying 1000 randomly generated software mappings for on-chip operations, along with 200 distinct hardware configurations for the Gemmini accelerator. This stochastic approach allows us to thoroughly explore the solution space and identify configurations that yield the most advantageous performance outcomes.

\begin{figure}[htp]
    \centering
    \includegraphics[width=9cm]{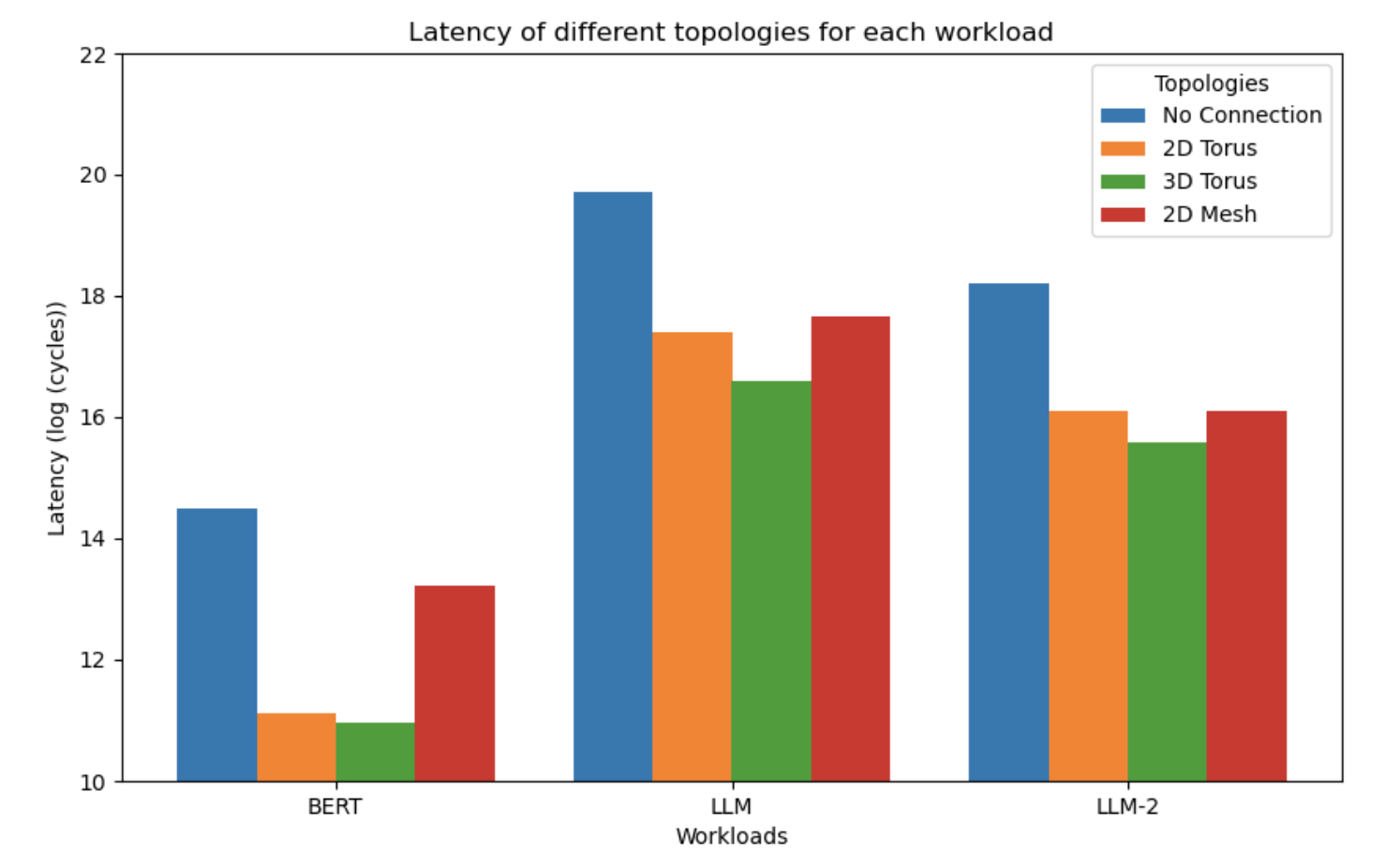}
    \caption{depicts the logarithmic latency, measured in cycles, of different inter-chip connection topologies (No Connection, 2D Torus, 3D Torus, and 2D Mesh) across three different workloads (BERT, LLM, and LLM-2).
}
    \label{fig:search1}
\end{figure}

In our initial exploration, we assess a variety of workloads on standard interconnection models to determine the optimal hardware topology. We evaluate the performance on a 2D Torus configuration with 8 chips ($2\times4$), a 3D Torus with 8 chips ($2\times2\times2$), a 2D Mesh with 8 chips, and a scenario with 8 chips no direct chip interconnection. This investigation serves to benchmark the efficiency of different topological structures and their impact on workload management, providing insights into how data flows and communication delays vary across these distinct network designs. Fig.\ref{fig:search1}

The study demonstrates that the presence of inter-chip connections substantially reduces latency across different workloads. Additionally, it reveals that while 2D Mesh, 2D Torus, and 3D Torus topologies offer improvements over systems with no chip interconnections, the differences in latency among these three interconnected topologies are marginal.

\begin{figure}[htp]
    \centering
    \includegraphics[width=9cm]{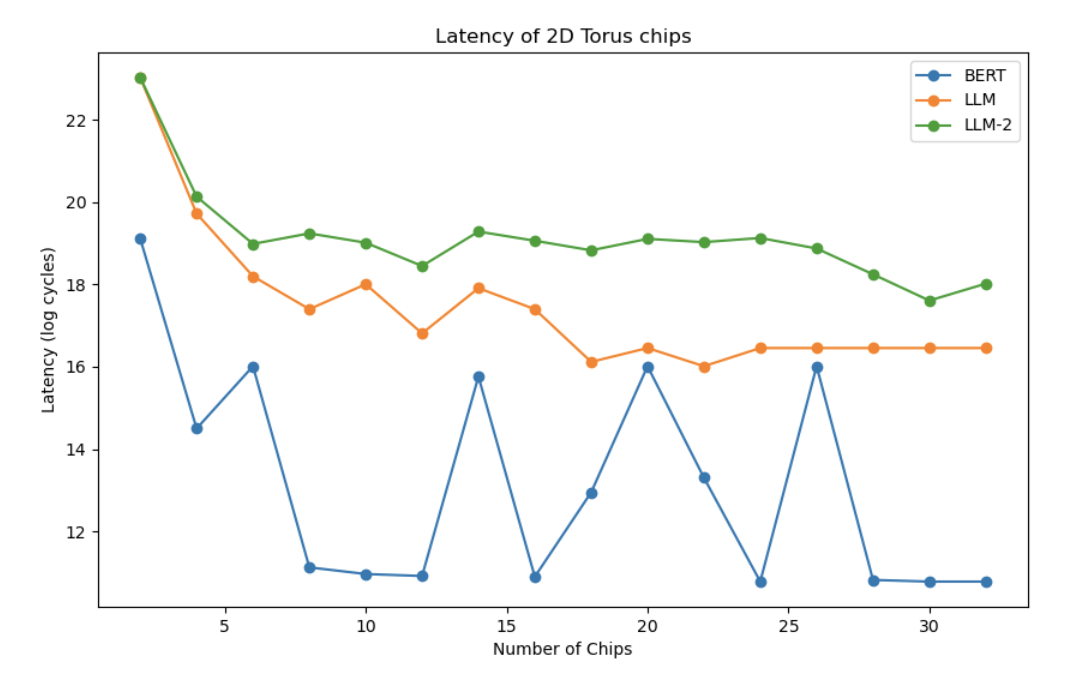}
    \caption{The graph shows the logarithmic latency in cycles for the BERT, LLM, and LLM-2 simulated workloads as the number of 2D Torus chips increases from 1 to 36.}
    \label{fig:search2}
\end{figure}

In our subsequent attempts of exploration, we hold the 2D Torus topology constant while varying the number of chips within the network. We explore all possible 2D torus topology in each iteration by fix the number of chips. This approach allows us to analyze the scalability of this particular configuration and understand how increasing or decreasing the number of chips affects overall system performance.  Fig.\ref{fig:search2}

In this study, initially, for the BERT workload, there is a steep decrease in latency as the number of chips increases, which then stabilizes, suggesting that beyond a certain point, adding more chips doesn't significantly improve latency. For the LLM workload, there's a consistent but slight downward trend, indicating a more linear relationship between the number of chips and latency reduction. The LLM-2 workload shows a more variable pattern where latency reduces, stabilizes, and even increases slightly before ending with a sharp decrease, indicating that the effect of adding more chips on latency might be more complex for this particular workload. The overall trend suggests that while increasing the number of chips generally leads to reduced latency, the degree of improvement depends on the specific characteristics and demands of the workload, and there might be an optimal number of chips beyond which the latency does not improve significantly.

The observed instability in latency reduction for the 2D Torus chips, particularly in the LLM-2 workload, can be attributed to the inherent limitations of the 2D Torus topology when accommodating certain numbers of chips. Since a 2D Torus topology necessitates a rectangular arrangement, certain chip counts—like 13—do not fit into a natural 2D grid and are forced into suboptimal configurations such as a 13x1 array. This results in inefficient topologies with longer inter-chip communication paths, leading to increased latency. Such irregular configurations disrupt the uniformity of the 2D Torus structure, thereby undermining its potential for latency optimization.

\begin{figure}[htp]
    \centering
    \includegraphics[width=9cm]{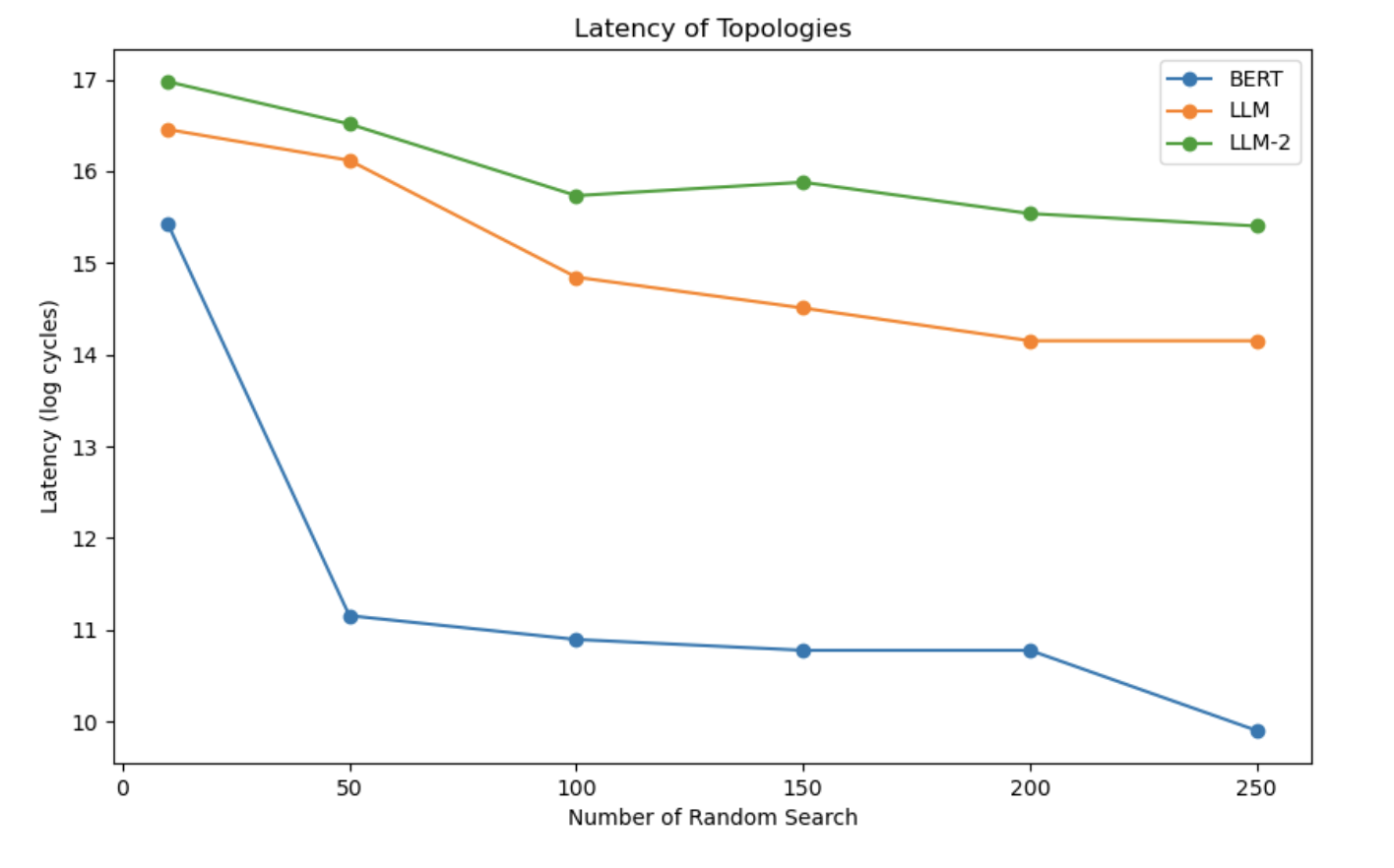}
    \caption{The graph displays the decrease in logarithmic latency (in cycles) for the BERT, LLM, and LLM-2 simulated workloads as the number of random topology searches increases up to 250.}
    \label{fig:search3}
\end{figure}

In the final phase of our exploration, we embrace a more exploratory and stochastic approach by randomly generating potential topologies. The only constraint of the topology is the number of chips is smaller than 50.This method allows us to investigate a wide array of interconnection patterns beyond conventional models. Fig.\ref{fig:search3}

The graph suggests that conducting a random search for optimal network topologies can significantly reduce latency for various workloads, with diminishing returns as the number of searches increases. The BERT workload shows the most substantial decrease in latency, indicating an efficient identification of optimal topologies. The LLM workload experiences a more gradual improvement, while the LLM-2 latency stabilizes quickly, suggesting an early discovery of an effective topology.

In assessing the optimal topology configurations, it was observed that topologies resembling a 3D Torus with a higher number of chips tend to be optimal. This preference likely arises from the incorporation of relative distances between chips into the latency formula for inter-chip connections, which favors configurations where these distances are minimized. The 3D Torus structure, inherently designed to reduce the average distance between nodes, consequently enhances communication efficiency and reduces latency, demonstrating the critical influence of physical layout on network performance.

\subsection{Real Hardware Evaluation}

\begin{figure}[htp]
    \centering
    \includegraphics[width=9cm]{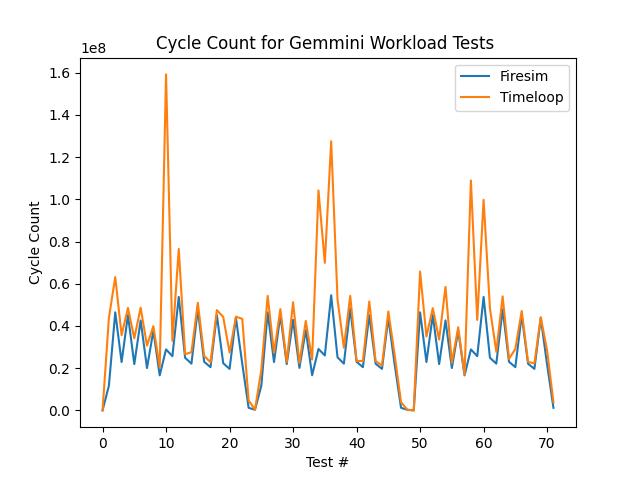}
    \caption{Comparing the cycle count for the exact same single-accelerator workload tests on Gemmini on Firesim vs Timeloop shows similar but slightly different cycle counts.}
\end{figure}

The reliance on Timeloop for simulations in the DeapSim platform is a critical aspect of the study. Timeloop, as a key simulation tool, plays an essential role in modeling the behavior of single-accelerator systems like Gemmini. However, a comparison of cycle counts between Timeloop and FireSim\cite{karandikar2018firesim}, another prominent simulation tool, reveals notable differences. Figure 12 shows these discrepancies, while not substantial, indicating that Timeloop may not fully account for system-level cycles, possibly leading to variations in accuracy.

Despite these differences, Timeloop's rapid simulation capabilities make it an invaluable tool for DSE. Its efficiency in simulating complex computational models is a significant advantage, especially when dealing with the extensive computations required by LLMs. One the other hand, the implementation of inter-chip communication simulation in a real hardware context indeed presents significant challenges, primarily due to the difficulty in accurately simulating wire delays. 

Future research could focus on enhancing Timeloop's model to include more detailed system-level dynamics, bridging the gap in accuracy observed in comparisons with tools like FireSim. Additionally, developing methods to more accurately simulate inter-chip communication in the absence of detailed RTL information could further refine the accuracy of multi-accelerator system simulations. These improvements would not only benefit DeapSim but also contribute broadly to the field of hardware-software co-design for advanced computational systems.

\section{Further Study}

It might be advantageous to explore to multi-accelerator simulation strategies by scaling up an entire end-to-end system. While our proposed framework can offer a reference, it's not apt for reporting cycle-exact metrics which can be limiting when prototyping hyperscale solutions. We propose using FireSim to extend the search space to include other components of serving an LLM system that would allow us to report cycle-exact metrics. Varying the client-facing communication protocol through RPCs or HTTP request can offer another dimension into the hardware-software co-design. 

Another limitation of our proposed framework involves the synchronous computation acrosss the accelerators. While incorporating this into simulation hosts a plethora of communication issues, we believe that asynchronous copy and compute is more indicate of workloads in the wild. Allow accelerators to begin computation asynchronously before another accelerator has loaded it's data would make scheduling into a NP-hard search problem, which is why we chose to omit it for the scope of this paper.

Moreover, one limitation of our proposed DeapSim architecture is its simplicity, featuring only a single accelerator per chip and a centralized CPU to orchestrate operations across the system. This design does not reflect the complexity of contemporary architectures like Google's TPU v4 or Meta's Zion, which deploy multiple accelerators on each chip to enhance parallel processing and computational throughput.

\section*{Acknowledgements}
 This is the class project for CS294-252 at University of California, Berkeley. Authors appreciate Professor Krste Asanovic and Sagar Karandikar for their invaluable guidance and support. We would also like to extend special thanks to Professor Sophia Shao for sponsoring the A-machine access in SLICE Lab and Charles Hong for Timeloop data.

%%%%%%% -- PAPER CONTENT ENDS -- %%%%%%%%

%%%%%%%%% -- BIB STYLE AND FILE -- %%%%%%%%
\bibliographystyle{IEEEtranS}
\bibliography{refs}
%%%%%%%%%%%%%%%%%%%%%%%%%%%%%%%%%%%%

\end{document}